\input{aipcheck}

\documentclass[
    ,final            
  ]
  {aipproc}

\layoutstyle{6x9}

\usepackage{epsfig}

\newcommand{\qw}{\ensuremath{\tilde{q}}}

\newcommand{\gw}{\ensuremath{\tilde{g}}}

\newcommand{\mqw}{\ensuremath{m_{\tilde{q}}}}
\newcommand{\mgw}{\ensuremath{m_{\tilde{g}}}}

\newcommand{\G}{\ensuremath{{g}}}
\newcommand{\Gt}{\ensuremath{\tilde{g}}}

\begin{document}

\title{Distinguishing Majorana and Dirac\\ Gluinos and Neutralinos}


\author{A.~Freitas}{
  address={Department of Physics and Astronomy, University of Pittsburgh, PA 15260, USA}
}

\begin{abstract}
While gluinos and neutralinos are Majorana fermions in the MSSM, they can be
Dirac fermion fields in extended supersymmetry models. The difference between the two
cases manifests itself in production and decay processes at colliders. In this
contribution, results are presented for how the Majorana or Dirac nature of
gluinos and neutralinos can be extracted from di-lepton signals at the LHC.
\end{abstract}

\maketitle


\section{Majorana and Dirac fermions in supersymmetry}

In the Minimal Supersymmetric Standard Model (MSSM), the fermionic partners of
the neutral, self-conjugate gauge bosons are self-conjugate Majorana fields with two degrees of
freedom each. For example, the relation between the gluon and gluino fields and
their charge conjugate fields is given by ${\G}^c  = - \, {\G}^T, 
  {\Gt}^c = - \, {\Gt}^T$,
where color indices have been suppressed.
Massive Majorana particles are known to mediate fermion-number violating 
production and decay processes, such as
\begin{equation}
q_L q_L \to {\tilde{q}}_L {\tilde{q}}_L,
\quad
q_R q_R \to {\tilde{q}}_R {\tilde{q}}_R,
\qquad
\tilde{q}_L \to q \,l^\pm \,\tilde{l}^\mp.
\end{equation}
These processes thus are central for experimentally testing the Majorana nature
of gluinos.

To study the characteristic differences between Majorana and Dirac fields, we
will focus on a well-defined framework where the gauge fields are embedded in
$N=2$ superfields \cite{n2}. Each $N=2$ gauge hypermultiplet contains one vector
field, two two-component fermion fields, and one complex scalar fields, see
Tab.~\ref{ftab}.
Depending on the structure of supersymmetry (SUSY) breaking, the two fermion components can
form two distinct Majorana fields or one Dirac field. The scalar component could
lead to interesting phenomenology on its own \cite{n2scal}, but will not be discussed
here. The two MSSM Higgs doublets can be joined into an $N=2$ chiral/anti-chiral
hypermultiplet, but the quark and lepton fields will be restricted to $N=1$
representations as in the MSSM for the purpose of this work.
The results presented here are based on Ref.~\cite{paper}. Some earlier work on
the observability of the Majorana nature may be found in Ref.~\cite{earlier},
and consequences for the density of cold dark matter have been elaborated in
Ref.~\cite{beletal}.

\begin{table}
\begin{tabular}{c@{\hspace{2em}}ccc}
\hline
\tablehead{1}{l}{c}{Group}  &  
\tablehead{1}{c}{c}{Spin 1}     & 
\tablehead{1}{c}{c}{Spin 1/2}   & 
\tablehead{1}{c}{c}{Spin 0}         \\
\hline
SU(3)  & $g$         & $\tilde{g}; {\tilde{g}}'$  & $\sigma_g$     \\
SU(2)  & $W^\pm,W^0$ & ${\tilde{W}}^\pm,{\tilde{W}}^0;
            {\tilde{W}}^{\prime\pm},{\tilde{W}}^{\prime 0}$ &
                                    ${\sigma}_W^\pm,{\sigma}_W^0$    \\
U(1)   & $B$         & $\tilde{B};{\tilde{B}'}$   & $\sigma_B$     \\
\hline
\end{tabular}
\caption{The N=2 gauge hyper-multiplets.}
\label{ftab}
\end{table}

\section{The SUSY-QCD sector}

In our setup, the two gluino components $\gw$ and $\gw'$ in  the gluon vector
hypermultiplet each have the usual kinetic terms, but only the standard gluino
interacts with matter:
\begin{equation}
    {\mathcal{L}}_{\rm gluino\ interactions} = g_s {\rm Tr}\,
    \bigl ( {\overline{\Gt} \gamma^{\mu} [{\G}_{\mu}, {\Gt}]} +
          {\overline{\Gt'} \gamma^{\mu} [{\G}_{\mu}, {\Gt}']} \bigr ) 
  - g_s \left[\,{\overline{q_L}  \gw \, \tilde{q}_L
                 - \overline{q_R}  \gw \, \tilde{q}_R}
                 + {\rm h.c.}\right]\,.
\end{equation}
The soft supersymmetry breaking mass terms are given by
\begin{equation}
    {\mathcal{L}}_{\rm soft\ gluino\ masses}
    =   - \frac{1}{2}
        \left[ M_3'\, {\rm Tr}(\overline{\Gt'}  {\Gt}')
        + M_3\, {\rm Tr}(\overline{\Gt}  \Gt)
        + M_3^D\, {\rm Tr}(\overline{\Gt'} \Gt +
                         \overline{\Gt}  \Gt' ) \right]
           \,,
\end{equation}
so they form a $2\times2$ matrix in the ${\gw,\gw'}$-space.
In the limit $M'_3 \to \infty$ the $\gw'$ decouples and the MSSM gluino sector is
recovered. On the other hand, if $M_3=M'_3 =0$ and $M_3^D \neq 0$, the mass matrix has
two degenerate eigenvalues. In this case, the two Majorana states are paired into
one Dirac field $\gw_D = \frac{1}{2}(1+\gamma_5)\gw +
\frac{1}{2}(1-\gamma_5)\gw'$ with Dirac mass $M_3^D$.

\begin{figure}
\epsfig{figure=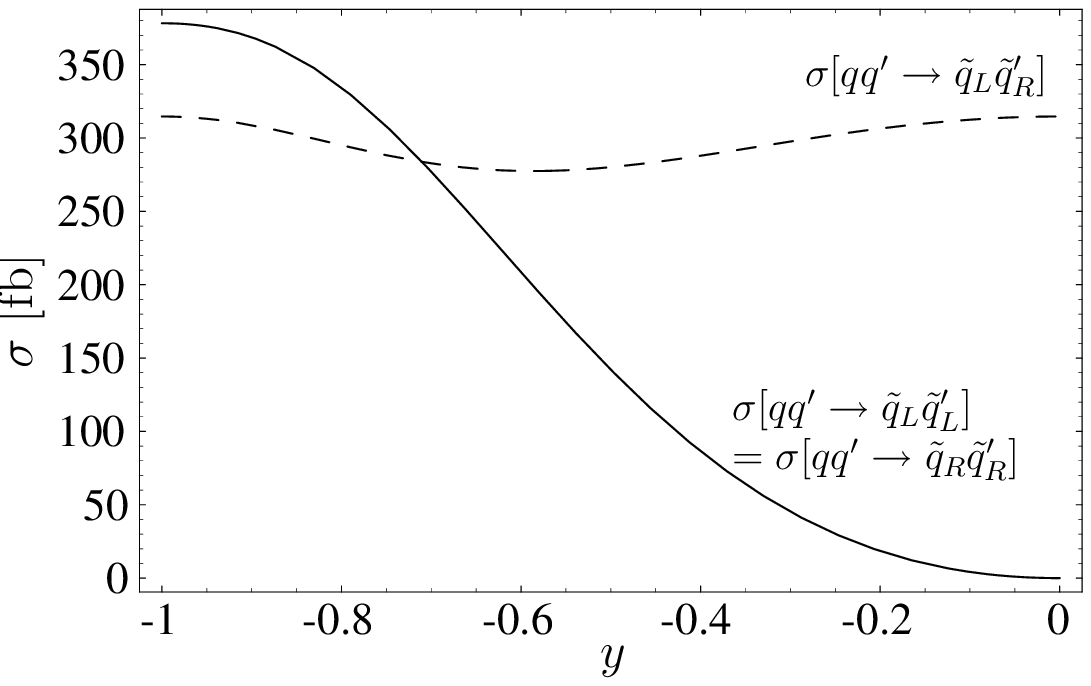, height=1.75in}\hspace{1em}
\epsfig{figure=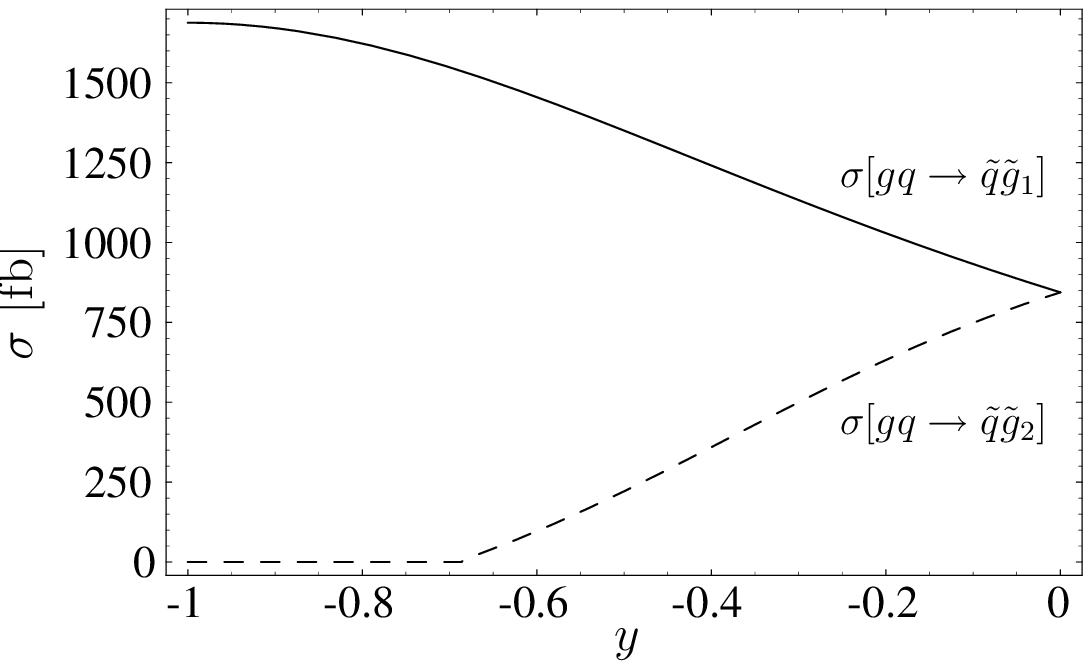, height=1.75in}
\caption{Partonic cross sections for $\qw\qw$ and $\qw\gw$ production at a
function of the Dirac/Majorana control parameter~$y$. The plot corresponds to a
fixed partonic center-of-mass energy $\sqrt{s}=2000$ GeV, and
$m_{\qw} = 500$ GeV and $m_{\gw_1} = 600$ GeV. }
\label{xsec}
\end{figure}

A smooth path can be defined that interpolates between the Majorana and
Dirac limits:
\begin{equation}
M'_3 = m_{\tilde{g}_1} \, \frac{y}{1+y}, \quad
M^D_3   = m_{{\gw}_1}, \quad
M_3 \,\,  = m_{{\gw}_1} M'_3 / (M_3' - m_{{\gw}_1})\,.
\end{equation}
As $y$ is varied between $-1$ and 0, one of the mass eigenvalues is kept fixed
at $m_{{\gw}_1}$, while the second eigenvalue changes from $\infty$ to
$m_{{\gw}_1}$. Therefore $y=-1,0$ correspond to the Majorana and Dirac limits,
respectively, while for any value in-between we obtain two Majorana gluino mass
eigenstates $\tilde{g}_{1,2}$ that are related to $\tilde{g}$ and $\tilde{g}'$ by a non-trivial
mixing matrix. A few examples for the $y$-dependence of partonic cross
sections for gluino and squark production are shown in Fig.~\ref{xsec}.\

Since, therefore, the ratio of gluino and squark production rates is different
in the Majorana and Dirac limits, this leads to observable effects for di-lepton
SUSY signatures at the LHC. In particular, assuming a standard scenario (such as
the SPS1a$'$ scenario \cite{sps}) with a
bino LSP, $\mgw > \mqw$, and the dominant decay chains
\begin{equation}
\tilde{q}_L \to q \, \tilde\chi^\pm_1 \to
q\,l^\pm \nu_l\tilde\chi^0_1, \qquad
\tilde{q}_R \to q \, \tilde\chi^0_1,
\end{equation}
the charge of the lepton is related to the charge of the L-squark. This results
in a net difference in the ratio of $l^+l^+$ and $l^-l^-$ rates between the
Majorana and Dirac case, see Tab.~\ref{dil}.
\begin{table}[b]
\begin{tabular}[t]{l@{\hspace{3em}}rr@{\hspace{3em}}rr@{\hspace{3em}}cc}
\hline
 & \tablehead{2}{c}{c}{Majorana}
 & \tablehead{2}{c}{c}{Dirac}
 & \tablehead{2}{c}{c}{\textit{N(l$^{\mbox{\bf +}}$l$^{\mbox{\bf +}}$)/
 				N(l$^{\mbox{\bf --}}$l$^{\mbox{\bf --}}$)}}\\
\cline{2-7}
 & $\sigma_{\rm tot}$ & $\sigma_{ll}$ after cuts &
          $\sigma_{\rm tot}$ & $\sigma_{ll}$ after cuts &
               Majorana       &    Dirac          \\
\hline
$\tilde{q}_L \tilde{q}_L^{(\prime)}$ &
    2.1 pb & 6.1 fb & 0 & 0 & 2.5  & $-$  \\
$\tilde{q}_L \tilde{q}_L^{(\prime)*}$ &
    1.4 pb & 3.1 fb & 1.4 pb & 3.1 fb & 1.4 & 1.4 \\
$\tilde{q}_L\gw_{(D)}^{\phantom{()}}$ &
    7.0 pb & 7.6 fb & 7.0 pb & 7.6 fb & 1.5 & 1.5 \\
$\gw_{(D)}^{\phantom{()}}\gw_{(D)}^{(c)}$ &
    3.2 pb & 1.4 fb & 7.0 pb & 3.2 fb & 1.0 & 1.0 \\
\hline
SM & 800 pb & $<$0.6 fb & 800 pb & $<$0.6 fb &
\multicolumn{2}{c}{1.0} \\
\hline
\end{tabular}
\caption{Signal and background cross sections in the SPS1a$'$ scenario \cite{sps},
before ($\sigma_{\rm tot}$) and after ($\sigma_{ll}$) including branching ratios and applying the
cuts of Ref.~\cite{yuk2}. The numbers always include also the charge conjugate
of the processes in the first column. A $pp$ center-of-mass energy of 14 TeV is
assumed.}
\label{dil}
\end{table}
After applying the cuts  of Ref.~\cite{yuk2} to
reduce the Standard Model (SM) background  and
by combining information of the total di-lepton rates and the jet
$p_\perp$ distributions, the two cases can be distinguished with a statistical
significance of 11 standard deviations, for an integrated luminosity of
300~fb$^{-1}$. As we have checked in Ref.~\cite{paper} this result is not
deteriorated significantly by including realistic uncertainties for the squark
and gluino masses, the missing higher-order corrections, and the parton
distribution functions.

\section{The electroweak sector}

The electroweak $N=2$ gauge multiplets can form Dirac neutralinos in a similar
fashion as explained for Dirac gluinos in the previous section. The matching of
the two Majorana neutralinos is straightforward if the mixing between gauginos
and higgsinos due to electroweak symmetry breaking can be neglected.

\begin{figure}
\begin{tabular}{c}
\epsfig{figure=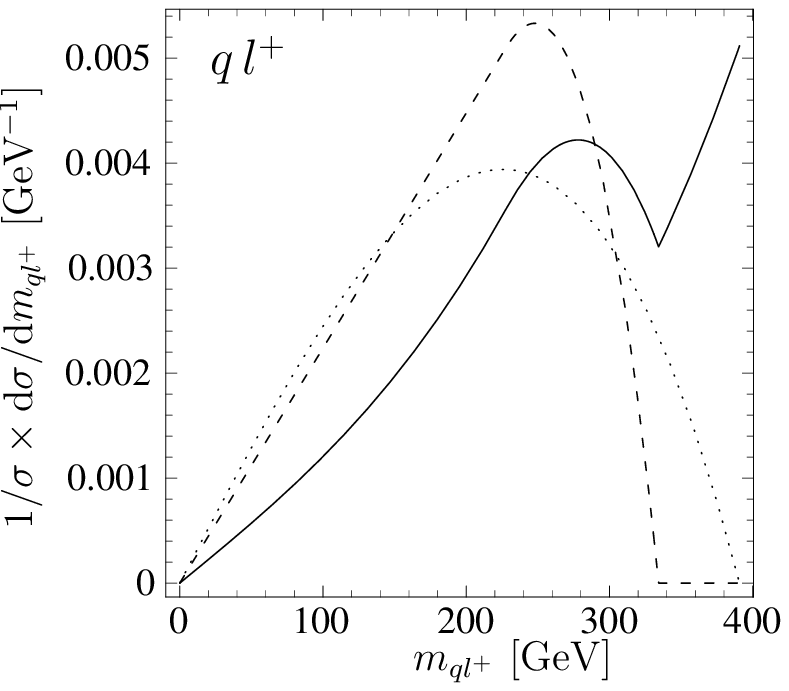, height=1.8in}
\hspace{-2mm}
\epsfig{figure=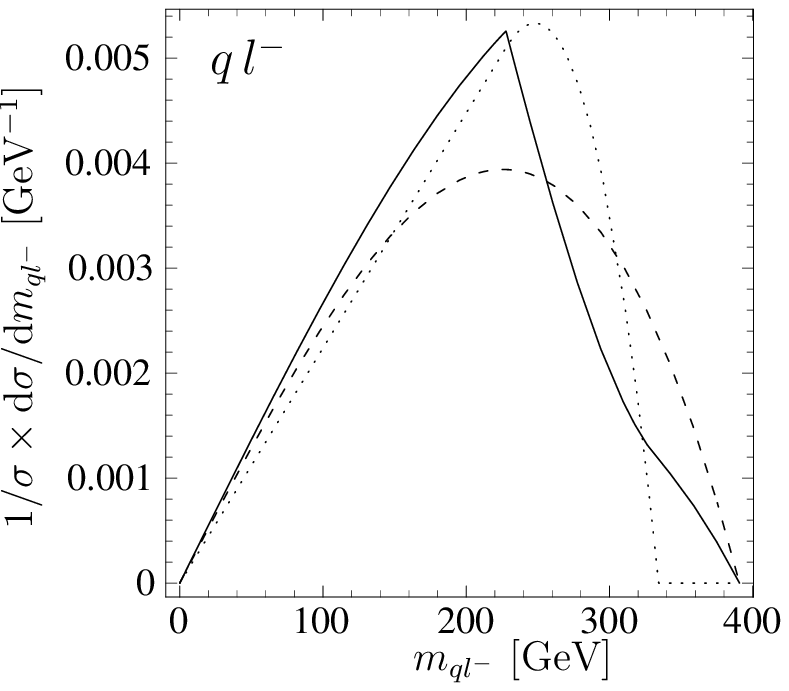, height=1.8in, bb=48 486 228 682, clip=true}
\hspace{-2mm}
\epsfig{figure=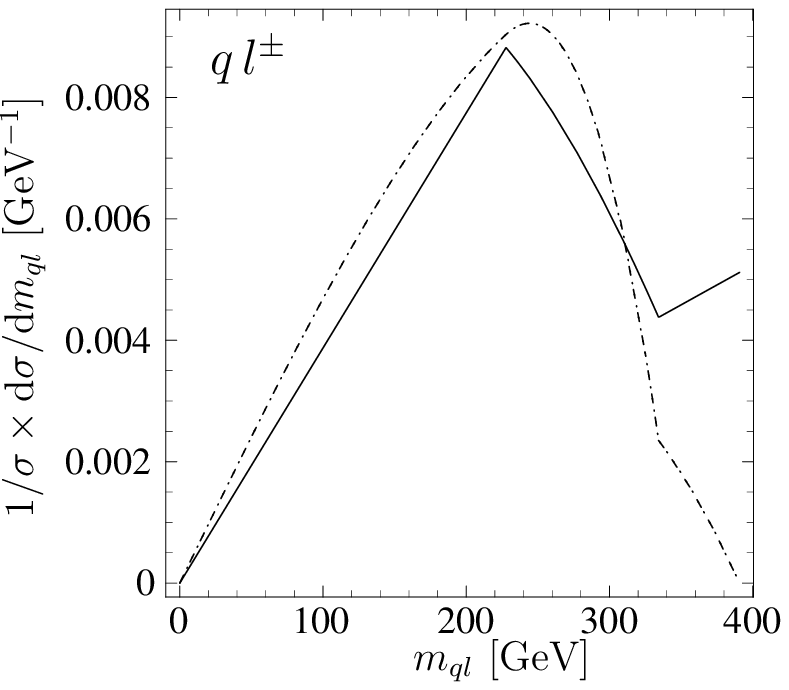, height=1.8in, bb=15 486 228 682, clip=true}
\\
\fbox{\parbox[t]{5.6in}{\AIPfigurecaptionheadfont
 ------
$\; \tilde{q}_L \to q \tilde\chi^0_2 
               \to q l^\mp_n l^\pm_f  \tilde\chi^0_1$
\hfill
-- -- --
$\; \tilde{q}_L \to q \tilde\chi^{c0}_{D2}
               \to q l^-_n l^+_f \tilde\chi^{c0}_{D1}$
\hfill
$\cdot\cdot\cdot\cdot\cdot$
$\; \tilde{q}_L^* \to \bar{q} \tilde\chi^0_{D2}
               \to \bar{q} l^+_n l^-_f \tilde\chi^0_{D1}$}} 
\end{tabular}
\caption{$ql$ invariant mass distributions for squark decay chains involving
Majorana or Dirac neutralinos. The masses have been taken from the SPS1a$'$
scenario \cite{sps}.}
\label{dist}
\end{figure}

The difference between the Dirac and Majorana theory leads to interesting
consequences for decay chains involving neutralinos at the LHC. Owing to its
self-conjugate nature, a Majorana
neutralino can decay into R-sleptons of either charge,
\begin{equation}
\tilde{q}_L \to q\, \tilde\chi^0_2 \to q\, l^\mp_n \tilde{l}_R^\pm
               \to q\, l^\mp_n \, l^\pm_f \, \tilde\chi^0_1\,.
\end{equation}
Here $l_n$ stands for the ``near'' lepton emitted in the first decay stage of
\pagebreak[0]
the neutralino, while $l_f$ denotes the ``far'' lepton stemming from the slepton
\pagebreak[0]
decay.
On the other hand, the corresponding Dirac neutralino ${\tilde\chi}^{0}_{D2}$ decays only
to negative R-sleptons, while its charge conjugate ${\tilde\chi}^{c0}_{D2}$ decays only
to positive R-sleptons:
\begin{equation}
   \tilde{q}_L \to q\, {\tilde\chi}^{c0}_{D2}
               \to q\, l^-_n\, \tilde l^+_R
               \to q\, l^-_n\, l^+_f\, {\tilde\chi}^{c0}_{D1}  , \qquad
   \tilde{q}_L^* \to \bar{q}\, \tilde\chi^0_{D2}
               \to \bar{q}\, l^+_n\, \tilde l^-_R
               \to \bar{q}\, l^+_n\, l^-_f\, \tilde\chi^0_{D1}\,.
\end{equation}
As a result, the $ql^\pm$ distributions arising from these decay chains are
markedly different for the Majorana and Dirac cases. This is shown in
Fig.~\ref{dist} for the SPS1a$'$ scenario.
The kink in the plots corresponds to the kinematic endpoint for the $ql_f$ system,
while the $ql_n$ endpoint is larger in this scenario and
coincides with the right edge of the plots.

\section{Summary}

SUSY extensions of the SM predict fermions in the
adjoint gauge group representations, which can be Majorana or Dirac fermions,
depending on the details of the model. Experimental distinction between the two
cases is thus of central interest for the physics program at future colliders.
Here it was shown how the Majorana or Dirac nature of gluinos can be determined
from SUSY production processes at the LHC, while neutralinos can be
analyzed in leptonic SUSY decay chains.





\bibliographystyle{aipproc}   

\end{document}